\documentclass[11pt,a4paper]{article}
\usepackage{jcappub}

\ifx\pdfoutput\undefined
\usepackage[dvips,bookmarks=false]{hyperref}	
\else
\usepackage{hyperref}	
\fi
\hypersetup{colorlinks,bookmarksopen,bookmarksnumbered,citecolor=blue,
linkcolor=black,pdfstartview=FitH,urlcolor=blue}

\usepackage{graphicx}
\usepackage{amsmath}
\usepackage{amssymb}

\usepackage{color}

\newcommand{\be}{\begin{equation}}
\newcommand{\ee}{\end{equation}}
\newcommand{\beq}{\begin{equation}}
\newcommand{\eeq}{\end{equation}}

\newcommand{\vect}[1]{\boldsymbol{\rm #1}}

\newcommand{\vv}{\textbf{v}}

\makeatletter
\renewcommand{\fnum@table}{\textbf{\tablename~\thetable}}
\renewcommand{\fnum@figure}{\textbf{\figurename~\thefigure}}
\makeatother

\title{Is the effect of the Sun's gravitational potential on dark matter particles observable?}

\def\mpi{Max-Planck-Institut f{\"u}r Kernphysik,\\ 
Saupfercheckweg 1, 69117 Heidelberg, Germany}
\def\sthlm{Oskar Klein Centre for Cosmoparticle Physics, 
Department of Physics,\\ 
Stockholm University, SE-10691 Stockholm, Sweden}

\author[a]{Nassim Bozorgnia}
\author[b]{and Thomas Schwetz}

\affiliation[a]{\mpi}
\affiliation[b]{\sthlm}

\emailAdd{bozorgnia@mpi-hd.mpg.de}
\emailAdd{schwetz@fysik.su.se}

\abstract{We consider the effect of the Sun's gravitational potential
  on the local phase space distribution of dark matter particles,
  focusing on its implication for the annual modulation signal in
  direct detection experiments. We perform a fit to the modulation
  signal observed in DAMA/LIBRA and show that the allowed region
  shrinks if Solar gravitational focusing (GF) is included compared to
  the one without GF. Furthermore, we consider a possible signal
  in a generic future direct detection experiment, irrespective of the
  DAMA/LIBRA signal. Even for scattering cross sections close to the
  current bound and a large exposure of a xenon target with 270 ton yr
  it will be hard to establish the presence of GF from data. In the
  region of dark matter masses below 40~GeV an annual modulation
  signal can be established for our assumed experimental setup,
  however GF is negligible for low masses. In the high mass region,
  where GF is more important, the significance of annual modulation
  itself is very low. We obtain similar results for lighter targets such as Ge and Ar.
   We comment also on inelastic scattering, noting
  that GF becomes somewhat more important for exothermic scattering
  compared to the elastic case.}


\begin{document}
\maketitle

\section{Introduction}
\label{sec:introduction}

The effect of the Sun's gravitational potential on the local phase
space distribution of Galactic dark matter (DM) particles and its
effect on DM direct detection experiments have been considered by a
number of authors \cite{Griest:1987vc, Sikivie:2002bj,
  Alenazi:2006wu}, see also \cite{1957MNRAS.117...50D,
  1967AJ.....72..219D, Patla:2013vza}. It turns out that the effect on
the total scattering rate of DM particles is very small. Recently, in
Ref.~\cite{Lee:2013wza}, it was pointed out, however, that Solar
gravitational focusing (GF) may have a sizable effect on the phase of
the annual modulation in the DM--nucleus scattering rate. Such a
modulation is induced by the motion of the Earth around the Sun, while
the Sun moves through the Galactic halo \cite{Drukier:1986tm,
  Freese:1987wu}. On top of this effect due to boosting the dark
matter velocity distribution into the Earth's rest frame, an
additional modulation is induced by the distortion of the DM phase
space density through the Sun's gravitational potential. The boost
effect has its minima and maxima around June and December, when the
Sun's and Earth's velocities add up or subtract. The GF induced
modulation will have minima and maxima when the Earth is located
behind or in front of the Sun, which happens in March and September,
respectively. Hence, the net-modulation will emerge as an interplay of
those two effects, which may significantly affect the phase of the
signal.

In section~\ref{sec:Annual modulation} below, after fixing the
notation, we are going to discuss those effects in terms of the
so-called halo integral, which captures the time dependence of the
signal in a particle physics as well as experimental configuration
independent way. In the following we present numerical studies of this
effect, investigating whether GF can be established from data, as well
as its impact on extracting DM parameters. As a first case study we
consider in section~\ref{sec:DAMA} the annual modulation signal
reported by the DAMA/LIBRA collaboration~\cite{Bernabei:2010mq} and
show that taking into account GF leads to a more constrained region in
the plane of DM mass and scattering cross section. However, the
relevant region in parameter space is highly excluded by data from
several other experiments. In particular, the LUX (Large Underground
Xenon) experiment~\cite{LUX:2013} has set the strongest limits on the
spin-independent elastic WIMP-nucleon cross section, with a minimum
upper limit of $7.6 \times 10^{-46}$ cm$^2$ on the cross section at
90\% CL for a DM mass of 33~GeV assuming the Standard Halo Model.
Therefore, we proceed in section~\ref{sec:Xe} by discarding the
DAMA/LIBRA signal and consider a hypothetical future experiment. To be
specific, we assume a large scale liquid xenon experiment, along the
lines discussed in \cite{Aprile:2012zx, XENONnT, Baudis:2012bc,
  Malling:2011va}. Our results indicate that even with a very large
exposure of several 100~ton~yr it will be difficult to establish an
annual modulation signal at high significance where GF has an
observable effect. In subsection~\ref{sec:light} we comment briefly on
light target nuclei. For most part of the work we assume elastic
spin-independent scattering; in subsection~\ref{sec:inelast}
we comment also on inelastic scattering. We summarize in
section~\ref{sec:summary}.

\section{Annual modulation with gravitational focusing}
\label{sec:Annual modulation}

\subsection{Notation}

The differential rate in events/keV/kg/day for a dark matter particle $\chi$ to scatter off a nucleus $(A,Z)$ and deposit the nuclear recoil energy $E_{nr}$ in the detector is given by
\beq \label{rate}
{R}(E_{nr},t) = \frac{\rho_\chi}{m_\chi} \frac{1}{m_A}\int_{v>v_{m}}d^3 v \frac{d\sigma_A}{d{E_{nr}}} v f_{\rm det}(\vect v, t),
\eeq
where $\rho_\chi$ is the local DM density, $m_A$
and $m_\chi$ are the nucleus and DM masses, $\sigma_A$ is the DM--nucleus
scattering cross section and $\vect v$ is the 3-vector relative velocity
between DM and the nucleus, while $v\equiv |\vect{v}|$. $f_{\rm det}(\vect v, t)$ is the DM velocity distribution in the detector reference frame. 
For elastic scattering, the minimal velocity $v_{m}$ required for a DM particle to
deposit a recoil energy $E_{nr}$ in the detector is given by
\beq
v_m=\sqrt{\frac{m_A E_{nr}}{2 {\mu_{\chi A}^2}}},
\label{eq:vm}
\eeq
where $\mu_{\chi A}$ is the reduced mass of the DM--nucleus system.

The time dependence of the differential event rate is due to
the velocity of the Earth with respect to the Sun, $\vect{v}_e(t)$,
which can be written as \cite{Gelmini:2000dm}
\beq\label{eq:ve}
\vect{v}_e(t) = v_e [\vect{e}_1 \sin\lambda(t) - \vect{e}_2\cos\lambda(t) ] \,,
\eeq
with $v_e=29.8$ km/s, and $\lambda(t)=2\pi(t-0.218)$ with $t$ in
units of 1 year and $t=0$ at January 1st, while $\vect e_1 =
(-0.0670,0.4927,-0.8676)$ and $\vect e_2 =(-0.9931,-0.1170,0.01032)$
are orthogonal unit vectors spanning the plane of the Earth's orbit.
We are using Galactic coordinates where $x$ points towards the
Galactic Center, $y$ in the direction of the Galactic rotation, and
$z$ towards the Galactic North, perpendicular to the disc. As shown in
\cite{Green:2003yh}, Eq.~\eqref{eq:ve} provides an excellent
approximation to describe the annual modulation signal. However,
corrections to Eq.~\eqref{eq:ve} become relevant for higher harmonics
of the time dependence. When discussing bi-annual modulation in
section~\ref{sec:time-dep} we do include the eccentricity of the orbit
following Refs.~\cite{McCabe:2013kea, Lee:2013xxa}.

Including the effect of GF, the DM density times the velocity distribution in
Eq.~\eqref{rate} is obtained by 
\beq\label{eq:rhochi}
\rho_\chi f_{\rm det}(\vv, t)=\rho_\infty \tilde f(\vv_\odot +\vv_\infty[\vv+\vv_e])\,.
\eeq
Here $\rho_\infty = 0.3$ GeV$/$cm$^3$ and $\tilde f(\vv)$ are the DM density and velocity distribution in the Galactic rest frame measured near the Solar System, but far away from the Sun, such that the Sun's gravitational potential is small, and $\vv_\odot \approx (10,233,7)$~km/s is the velocity of the Sun with respect to the Galaxy. The function $\vv_\infty[\vv]$ relates the velocity $\vv_\infty$ of a DM particle relative to the Sun far away from the Sun's gravitational potential to the particle's velocity $\vv$ at the detector~\cite{Alenazi:2006wu}:
\beq\label{eq:vinf}
\vv_\infty[\vv]=\frac{v_\infty^2 \vv+v_\infty u_{\rm esc}^{2} \hat {\bf r}_s/2 -v_\infty \vv (\vv\cdot \hat {\bf r}_s) }{v_\infty^2+ u_{\rm esc}^{2}/2-v_\infty (\vv\cdot \hat {\bf r}_s) } \,,
\eeq
with $v_\infty^2=v^2-u_{\rm esc}^2$, where $u_{\rm esc}=\sqrt{2G M_{\odot}/r_{\rm A.U.}}\simeq 40$~km/s is the escape velocity from the Sun near the Earth's orbit, and is of ${\mathcal O}(v_e)$, while $\hat {\bf r}_s$ is the unit vector pointing in the direction of the Earth from the center of the Solar System, and is given by
\beq\label{eq:rs}
\hat {\bf r}_s = - [\vect{e}_1 \cos\lambda(t) + \vect{e}_2 \sin\lambda(t) ] \,.
\eeq
If GF is neglected, the velocity distribution in the detector's rest
frame is obtained just by a boost from the Galactic rest frame:
$f_{\rm det}(\vv, t)= \tilde f(\vv + \vv_\odot + \vv_e)$ and $\rho_\chi = \rho_\infty$.
In this paper we assume the so-called Standard Halo Model 
with a truncated Maxwellian velocity distribution,
\beq\label{eq:Maxwellian}
\tilde f(\vv) = \frac{1}{N} \left[\exp \left( -\frac{\vv^2}{ \bar{v}^2} \right) - \exp \left( -\frac{ v_{\rm esc}^2}{ \bar{v}^2} \right)   \right] \Theta (v_{\rm esc} - v) \,,
\eeq
where $N$ is a normalization constant and we adopt the fiducial values 
$\bar{v}=220$ km/s and $v_{\rm esc}=544$ km/s. 

We focus here on spin-independent elastic scattering and assume that
DM couples with the same strength to protons and neutrons. In this
case the differential cross section which enters in Eq.~\eqref{rate}
is
\begin{align}
  \frac{d\sigma_A}{dE_{nr}} = \frac{m_A A^2}{2\mu_{\chi p}^2 v^2} {\sigma_{\rm SI}} F^2(E_{nr}) \,, 
  \label{eq:dsigmadE}
\end{align}
where $\sigma_{\rm SI}$ is the spin-independent DM--nucleon scattering cross section, $\mu_{\chi p}$ is the reduced mass of the DM--nucleon system, and $F(E_{nr})$ is a form factor. 

Defining the halo integral as
\beq\label{eq:eta} 
\eta(v_m, t) \equiv \int_{v > v_m} d^3 v \frac{\tilde{f}(\vv_\odot +\vv_\infty[\vv+\vv_e], t)}{v} \,,
\eeq
the event rate is given by
\beq\label{eq:Rgamma}
R(E_{nr}, t) = C \, F^2(E_{nr}) \, \eta(v_m, t), 
\qquad\text{with}\qquad
C = \frac{\rho_\infty A^2 \sigma_{\rm SI}}{2 m_\chi \mu_{\chi p}^2}.
\eeq
The coefficient $C$ contains the particle physics dependence, while
$\eta(v_m,t)$ parametrizes the astrophysics. The number of events in an
energy interval $[E_1, E_2]$ and at a given time $t$ can be written as
\beq\label{eq:N} 
N_{[E_1,E_2]} (t) = MT \int_{0}^{\infty} dE_{nr}
G_{[E_1,E_2]} (E_{nr}) R(E_{nr}, t) \,, 
\eeq 
where $MT$ is the exposure of the experiment in units of kg~day, and
$G_{[E_1,E_2]} (E_{nr})$ is the detector response function which
includes the detection efficiencies and energy
resolution.\footnote{$N_{[E_1,E_2]} (t)$ in Eq.~\eqref{eq:N} should be
  considered as time-differential rate. However, we take $t$ to be a
  dimensionless variable in $[0,1]$, indicating the time of the year,
  whereas the total time of the exposure is included in $T$. The
  total number of events in the energy interval 
  $[E_1,E_2]$ is obtained by $\int_0^1 dt \, N_{[E_1,E_2]}(t)$.}  The
response function may be non-zero outside the interval $[E_1, E_2]$
due to the energy resolution.  The annual modulation signal in a given
energy interval can be computed by subtracting the time averaged
events from the total number of events in that energy interval. We
have
\beq\label{eq:A} 
A_{[E_1,E_2]} (t) = \frac{N_{[E_1,E_2]} (t) -
  \langle N_{[E_1,E_2]} (t)\rangle_t}{E_2 - E_1} \,, 
\eeq 
where $\langle N_{[E_1,E_2]}(t)\rangle_t$ is 
the number of events averaged over one year in the given
energy bin. The units are events/keV.

\subsection{Time dependence of the halo integral}
\label{sec:time-dep}

\begin{figure}
\begin{center}
 \includegraphics[height=0.45\textwidth]{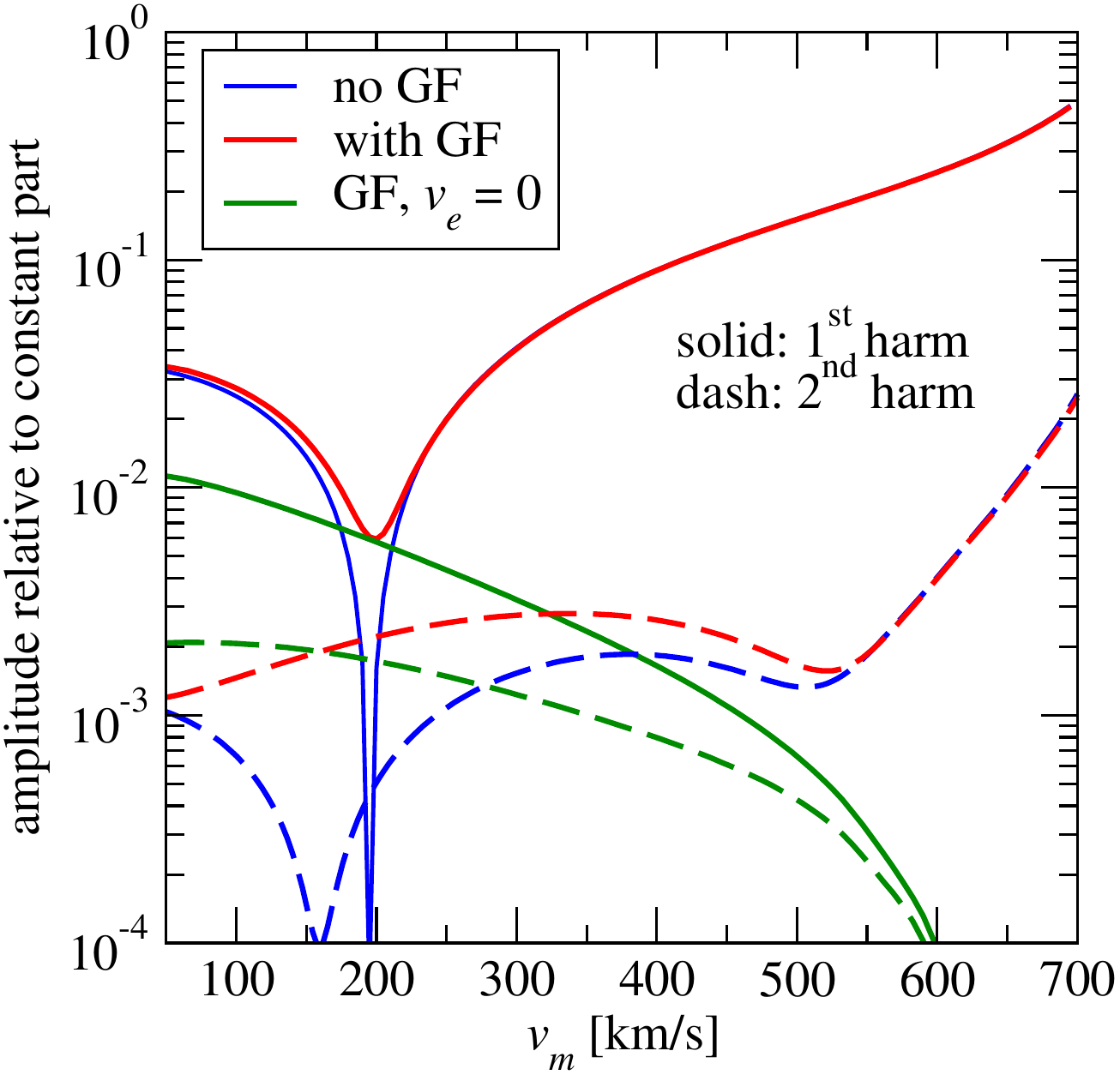} \quad
 \includegraphics[height=0.45\textwidth]{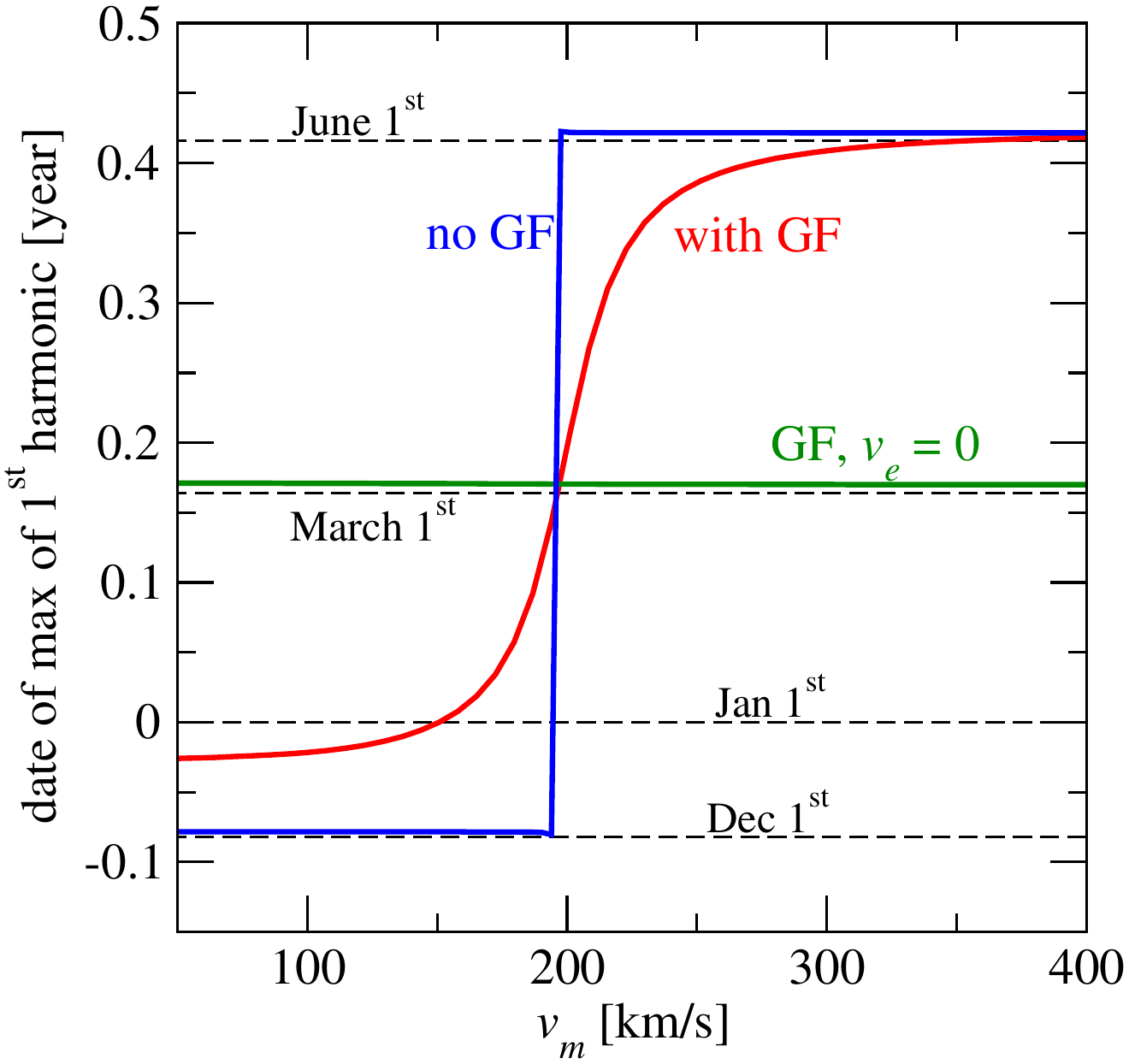}
\caption{\label{fig:eta} Time variation of the halo integral
  $\eta(v_m, t)$ as defined in Eq.~\eqref{eq:eta}. We assume a
  Maxwellian velocity distribution and perform a Fourier analysis of
  $\eta(v_m, t)$ for fixed minimal velocity $v_m$. The left plot shows
  the amplitudes of the first (solid) and second (dashed) harmonics
  relative to the time independent component. In the right panel we
  show the date of the maximum of the first harmonic as a function of
  $v_m$. In both panels we show the cases of neglecting Solar GF
  (blue), taking GF into account (red), and the effect of GF for
  setting the Earth's velocity relative to the Sun to zero (green).}
\end{center}
\end{figure}

Let us now discuss the time dependence of the event rate in a direct
detection experiment, which according to Eq.~\eqref{eq:Rgamma} is
determined from the time dependence of the halo integral defined in
Eq.~\eqref{eq:eta}. Note that the time variation of $\eta(v_m, t)$ is
independent of particle physics, in particular independent of the DM
mass. Assuming the Maxwellian, Eq.~\eqref{eq:Maxwellian}, we calculate
$\eta(v_m,t)$ numerically. For a fixed minimal velocity $v_m$ we
perform a Fourier decomposition in $t$. In the left panel of
Fig.~\ref{fig:eta} we compare the size of the amplitude of the first
harmonic (annual modulation) as well as the second harmonic (bi-annual
modulation) to the time-averaged component (zeroth order Fourier
coefficient). The right panel shows the date of the maximum of the
first harmonic. In this subsection we do not use Eqs.~\eqref{eq:ve}
and \eqref{eq:rs} to describe $\vect{v}_e(t)$ and $\hat {\bf r}_s(t)$,
but instead use the expressions given in the appendix of
Ref.~\cite{Lee:2013xxa} (see also \cite{McCabe:2013kea} for equivalent
results) including the eccentricity of the Earth's orbit. This is
relevant for the second harmonic. We have verified that for the first
harmonic corrections due to the eccentricity are negligible and
Eqs.~\eqref{eq:ve} and \eqref{eq:rs} are sufficient.

The blue curves in the figure correspond to the traditional case of
neglecting GF. The first harmonic shows the characteristic phase flip
close to $v_m\approx 200$~km/s, where the amplitude goes to zero and
below (above) that value of $v_m$ the maximum is at the beginning of
December (June). The second harmonic without GF has two phase flips
\cite{Chang:2011eb}, one around 150~km/s and another one around
500~km/s. Note that for the second phase flip the amplitude does not go 
to zero. The reason is related to the effect of the Earth's orbit
eccentricity, which makes the maximum shift smoothly by $\pi/2$ but
the amplitude remains non-zero. Once GF is included (red curves) we can
make the following observations: $(i)$ the amplitude of the first
harmonic is hardly affected by GF, the only exception being close to
the phase flip, with the amplitude never going to zero; $(ii)$ there
is a significant distortion of the maximum of the first harmonic
around and below the phase flip (see right panel). The date of the
maximum moves smoothly from around December 20 at very low $v_m$ to
beginning of June for large $v_m$.  Those results are in agreement
with \cite{Lee:2013wza}; and ($iii$) the amplitude of the second
harmonic is significantly affected by GF, leading even to the
disappearance of the phase flip at low $v_m$.

For illustration purpose we show in Fig.~\ref{fig:eta} also the effect
of GF but assuming that the Earth is at rest with respect to the Sun
(green curves). This corresponds to the hypothetical situation of
considering fixed positions of the Earth in turn and comparing the event
rates at the different locations of the Earth around the Sun. This procedure
isolates the effect of GF and removes the time dependence from the
velocity boost. Technically what we do is to set $v_e = 0$ but keep
the time dependence of $\hat {\bf r}_s$ in Eq.~\eqref{eq:vinf}. By
comparing the blue and green curves in the left panel we see the
relative importance of the time dependence induced by the velocity
boost and GF. While for the first harmonic GF is always subdominant
(except at the phase flip), for the second harmonic it is dominating
for $v_m \lesssim 300$~km/s. The right panel shows that the first
harmonic induced by GF only (green curve) always peaks at the beginning of March,
independent of $v_m$.

We do not discuss further the second harmonic, since
this will be very hard to observe in the foreseeable future. Focusing
on the first harmonic, we conclude that the main effect of GF is the
modification of the phase, especially for $v_m \lesssim
250$~km/s~\cite{Lee:2013wza}. In the rest of the paper we will perform
numerical studies of the importance (statistical significance) of this
effect in (semi)realistic experimental situations.

\section{Gravitational focusing and the DAMA/LIBRA signal}
\label{sec:DAMA}

The DAMA/LIBRA collaboration \cite{Bernabei:2010mq} reports a highly
significant annual modulation signal using an NaI target, which can be
interpreted in terms of elastic spin-independent DM scattering with DM
masses of around 10 or 80~GeV, depending on whether scattering happens
on the sodium or iodine nucleus. Despite the fact that the required
cross sections are excluded by a number of other experiments, we
consider their data as a case study in order to investigate the effect
of GF for extracting DM parameters. Ideally one would perform a fit to
the data using time as well as energy information. Unfortunately this
information is not public, since data are presented either binned in
time or in energy (but not showing the full 2-dimensional
information). Since the main effect of GF is a distortion of the time
behavior of the signal we are using here time binned data, extracted
from the top panel of Fig.~1 in Ref.~\cite{Bernabei:2010mq}. This
figure shows the time variation of the count rate in the $[2,4]$ keVee
energy interval corresponding to a 0.87~ton~yr exposure during about 5.5
years divided into 43 bins. In this energy interval, the modulation
signal is largest. We use $q_{\rm Na}=0.3$ and $q_{\rm I}=0.09$ for
the quenching factors of Na and I, respectively~\cite{Bernabei:1996},
and assume a Gaussian energy resolution with an energy dependent width
as presented in Ref.~\cite{Fairbairn:2008gz}.

To fit the DAMA data, we construct a $\chi^2$ function
\be
\chi^2_{\rm DAMA} (m_\chi, \sigma_{\rm SI}) = \sum_{i=1}^{43} \left(\frac{A_i^{\rm pred} (m_\chi, \sigma_{\rm SI}) - A_i^{\rm obs}}{\sigma_i} \right)^2,
\label{eq: DAMAchisq}
\ee
where $A_i^{\rm obs}$ and $\sigma_i$ are the experimental data points and their errors, respectively, from the top panel of Fig.~1 in Ref.~\cite{Bernabei:2010mq}. The sum is over the 43 time bins. The best fit point can be found by minimizing Eq.~(\ref{eq: DAMAchisq}) with respect to the WIMP mass $m_\chi$, and cross section $\sigma_{\rm SI}$. The allowed regions in the mass -- cross section plane at a given CL are obtained by looking for contours $\chi^2 (m_\chi, \sigma_{\rm SI}) =\chi^2_{\rm min} + \Delta \chi^2 ({\rm CL})$, where $\Delta \chi^2 ({\rm CL})$ is evaluated for 2 degrees of freedom (dof), e.g., $\Delta \chi^2 (90\%) = 4.6$.

\begin{figure}[t]
\begin{center}
 \includegraphics[width=0.46\textwidth]{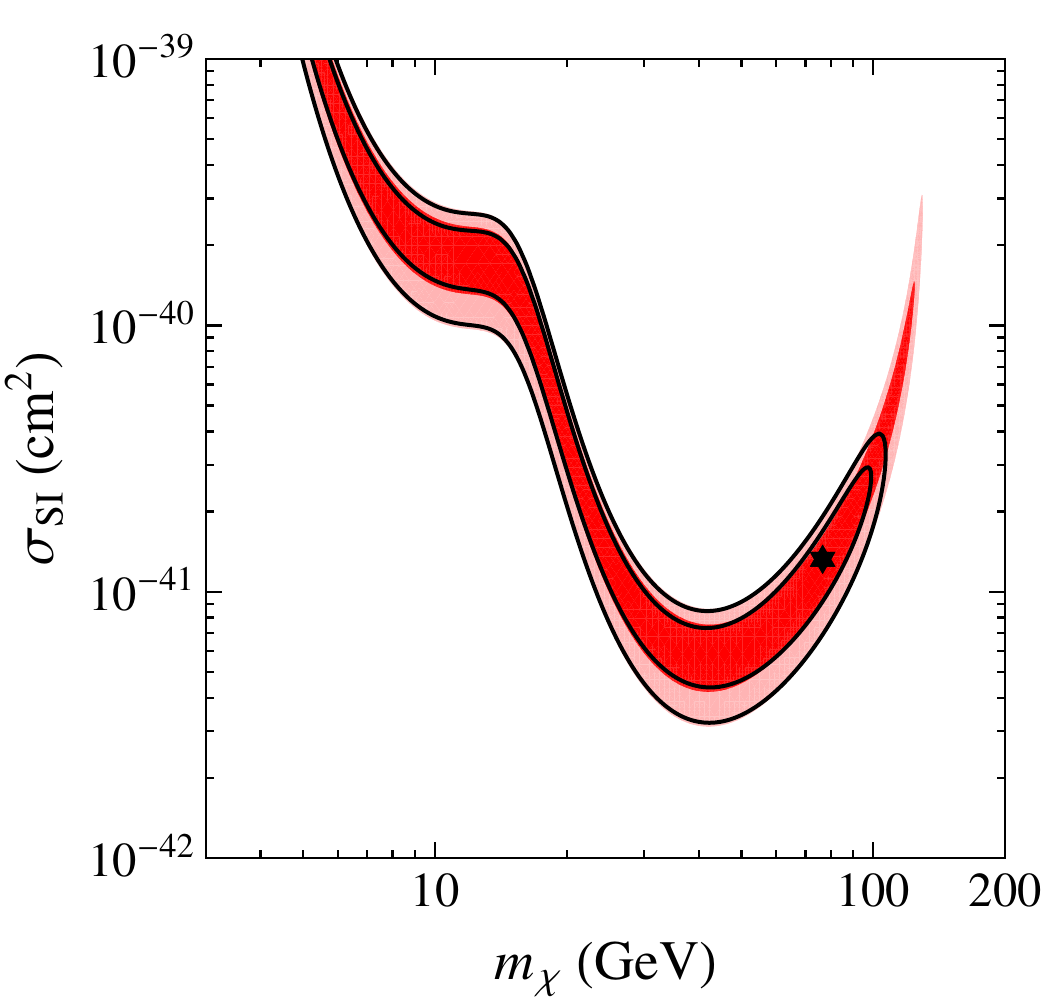}\hspace{5pt}
 \includegraphics[width=0.48\textwidth]{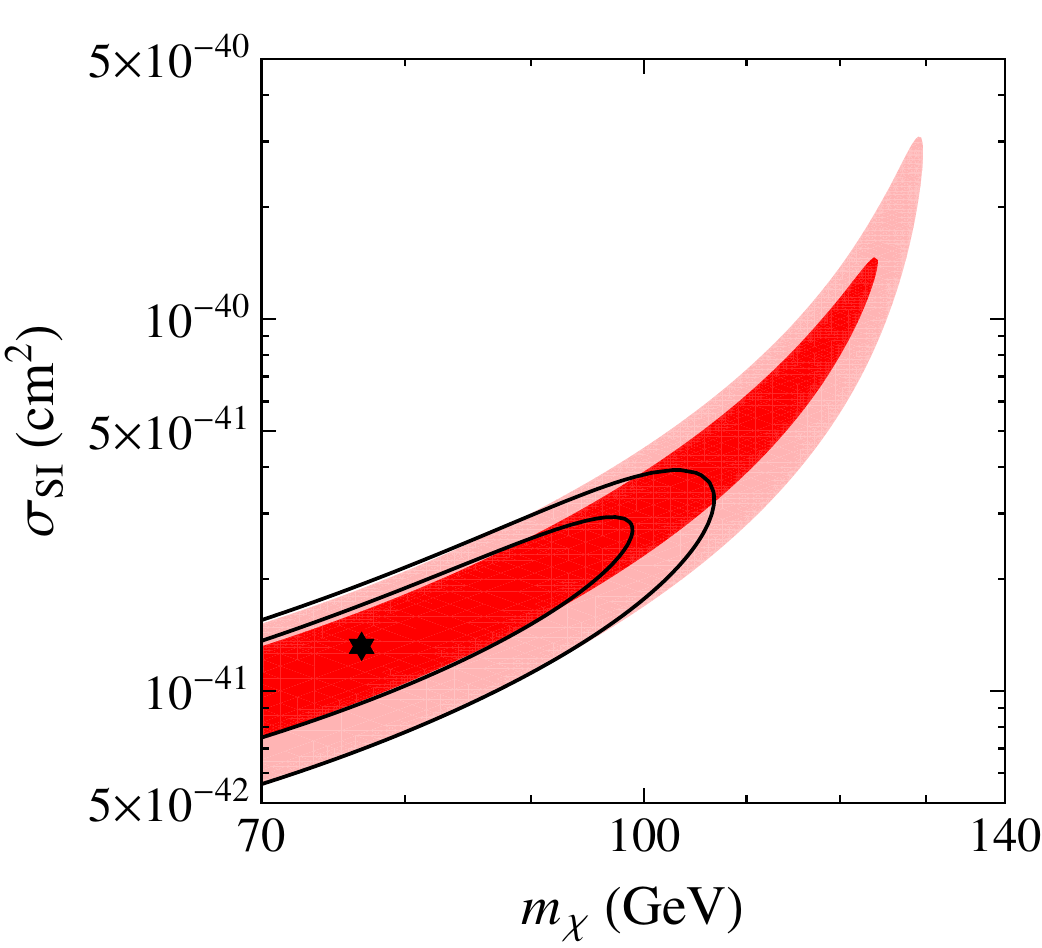}
\caption{\label{fig:DAMA} The preferred region of DAMA at 90\% CL and 3$\sigma$ with (black contours) and without (dark and light red regions) GF using time binned data in the single energy interval of 2 to 4~keVee. The left panel shows the result for a large range of dark matter masses, while the right panel zooms on the mass region where GF is important. The black star shows the best fit location with GF.}
\end{center}
\end{figure}

Fig.~\ref{fig:DAMA} shows the allowed region of DAMA at 90\% CL and
3$\sigma$ in the cross section and mass plane assuming $A_i^{\rm
  pred}$ is with (black contours) or without (dark and light red
regions) GF. With GF, $\chi^2_{\rm min} = 40.6$ for $m_\chi=76.9$ GeV
and $\sigma_{\rm SI}=1.3 \times 10^{-41}$ cm$^2$. The minimum is shown
with a star in Fig.~\ref{fig:DAMA}. Without GF, 
$\chi^2_{\rm min} \approx 42.2$ is practically degenerate along the
allowed strip in parameter space visible in the figure between DM
masses from $\sim5$~GeV to more than 100~GeV.
  
The $\chi^2_{\rm min}$ values in both cases are of order of the number
of degrees of freedom (43 data points minus 2 fitted parameters),
indicating a good fit without as well as with GF. However, one can see
from Fig.~\ref{fig:DAMA} that GF makes the preferred DAMA region
smaller, excluding the region of large DM masses. According to
Eq.~\eqref{eq:vm}, large masses correspond to small values of $v_m$,
and in this region the phase shift induced by GF becomes relevant,
leading to a disagreement with the data. In contrast, the region
$\lesssim 80$~GeV is practically unaffected by GF, which holds in
particular also for the $m_\chi \sim10$~GeV solution from scattering
on sodium. For such light DM masses, $v_m$ becomes large and only the
high velocity tail of the DM distribution is probed, where GF is
negligible, c.f., Fig.~\ref{fig:eta}.

Let us comment on the fact that the allowed region in
Fig.~\ref{fig:DAMA} appears as a degenerate band, in contrast to the
familiar two isolated regions (see \cite{Bozorgnia:2013pua} for a
recent example). As explained above, the fit on which
Fig.~\ref{fig:DAMA} is based on uses only limited information on the
energy dependence of the signal, since it uses time binned data in a
single energy interval. However, the energy spectrum is quite powerful
to constrain the allowed region \cite{Fairbairn:2008gz, Chang:2008xa}
and this information is missed in the current analysis. Ideally time
and energy information should be included simultaneously. In the
absence of this information we present an alternative fit to the DAMA data
using information from Fig.~9 of
Ref.~\cite{Bernabei:2010mq}\footnote{We thank the authors of
  Ref.~\cite{Lee:2013wza} for mentioning this possibility to us.}. There
the result of fitting a cosine function to the total exposure
(1.17~ton~yr, DAMA/NaI and DAMA/LIBRA) is presented, showing in
different energy bins the amplitude of the cosine, $Y_m$ (upper
panel), and the time of the maximum, $t^*$ (lower panel). Neglecting a
possible correlation between $Y_m$ and $t^*$ (as justified by Fig.~7
of \cite{Bernabei:2010mq}) we can use this to construct a $\chi^2$ function as
\be
\chi^2_{\rm DAMA'} (m_\chi, \sigma_{\rm SI}) = \sum_{i=1}^{7} 
\left(\frac{Y_{m,i}^{\rm pred} (m_\chi, \sigma_{\rm SI}) - Y_{m,i}^{\rm obs}}{\sigma_i^{Y_m}} \right)^2
+\sum_{j=1}^{4} \left(\frac{t_j^{*,\rm pred} (m_\chi, \sigma_{\rm SI}) - t_j^{*,\rm obs}}{\sigma_j^{t^*}} \right)^2 .
\label{eq:DAMAchisq-alt}
\ee
For $Y_m$ we use the 6 bins from 2 to 8~keVee and combine the
remaining bins from 8 to 20~keVee, where the modulation is consistent
with zero, into one single bin. For $t^*$ we use the 4 bins from 2 to
6~keVee, and ignore the data points above 6~keVee, where $t^*$ is
undetermined. The data points $Y_{m,i}^{\rm obs}, t_j^{*,\rm obs}$ and
the corresponding $1\sigma$ errors $\sigma_i^{Y_m}, \sigma_j^{t^*}$
are read off from Fig.~9 of Ref.~\cite{Bernabei:2010mq}.

 \begin{figure}[t]
\begin{center}
 \includegraphics[width=0.46\textwidth]{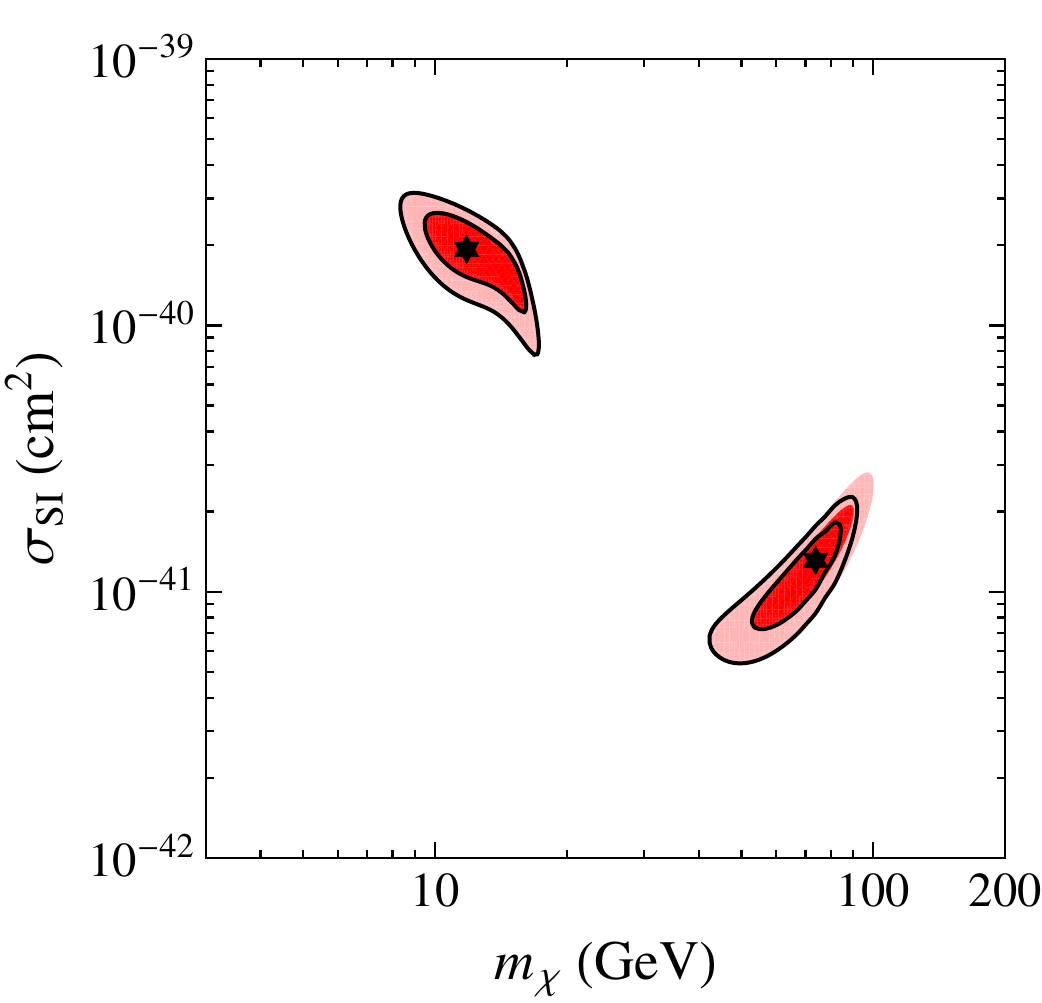}\hspace{5pt}
 \includegraphics[width=0.48\textwidth]{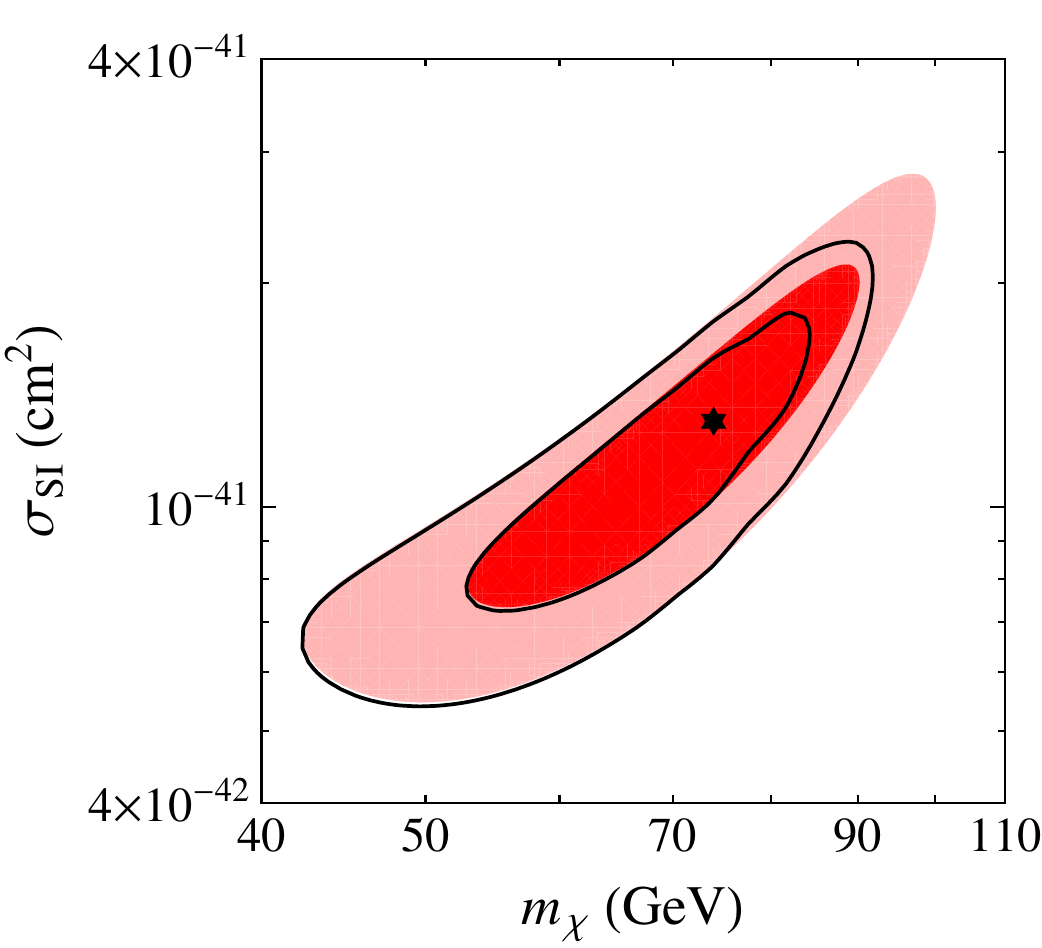}
\caption{\label{fig:DAMA-alt} The preferred region of DAMA at 90\% CL and 3$\sigma$ with (black contours) and without (dark and light red regions) GF using energy binned data on the amplitude and the phase of a cosine function fit to DAMA data. The left panel shows the result for a large range of dark matter masses, while the right panel zooms on the mass region where GF is important. The black star shows the best fit location with GF.}
\end{center}
\end{figure}

The results of this fit are shown in Fig.~\ref{fig:DAMA-alt}. Thanks
to the energy spectrum information we obtain the two islands
corresponding to scattering on sodium (low mass) and iodine (high
mass). Again the effect of GF is to constrain more the high-mass edge
of the iodine region. The difference of the regions with and without
GF is smaller than in Fig.~\ref{fig:DAMA}, because due to the energy
information the region without GF is already more constrained towards
high DM masses.

As mentioned above, the DM explanation of DAMA is in strong tension
with exclusion limits from other experiments. In particular, the
region around 80~GeV is excluded by the most recent limit from
LUX~\cite{LUX:2013} by around 4 orders of magnitude. In the following
we are going to discard the DAMA signal and discuss annual modulation
and GF in the context of a hypothetical future large scale direct
detection experiment.

\section{Annual modulation and GF in a future large scale experiment}
\label{sec:Xe}

\subsection{Simulation details and event numbers}
\label{sec:exp-details}

Let us assume that dark matter is just around the corner, slightly
below the current best limit from the LUX experiment~\cite{LUX:2013}
and investigate how the signal of GF would look like in a future large
scale direct detection experiment. To be specific, we assume a liquid
xenon detector, inspired by upgrade plans of the
XENON~\cite{Aprile:2012zx, XENONnT} and LUX~\cite{Malling:2011va}
collaborations, and the DARWIN consortium~\cite{Baudis:2012bc}. DARWIN
is supposed to be the ``ultimate'' direct detection experiment, and
exposures of order 10~ton~yr are considered. As we will see in the
following, it will be extremely difficult to establish an annual
modulation signal, given the current constraints on the scattering
cross section. Therefore, we are going to be very aggressive and
assume an exposure of $10^8$~kg~day~$\approx 270$~ton~yr. This
corresponds roughly to a factor $10^4$ larger than the current LUX
exposure, and could be achieved for instance by a $\sim 10$~yr
exposure of a hypothetical $\sim 30$~ton detector. We stress that such a
huge exposure goes beyond the currently discussed options. However, as
we will see, even with those (probably unrealistic) assumptions it
will be hard to establish statistically significant signals. Our
results are easily scalable to any exposure.

Further assumptions of our simulated setup are motivated by the LUX
and XENON100 analyses. For the detection efficiency we multiply
together the blue and dotted green curves in Fig.~1 of
Ref.~\cite{Aprile:2012nq}. Those curves are energy dependent, and give
the combined cut acceptance for the XENON100 detector, corresponding
to a hard discrimination cut used for the maximum gap method
analysis. The final efficiency is roughly 30\% for a large part of the
energy interval we consider; at 3~keV, the efficiency is 8\%, but
reaches a value of 20\% already at 4.5~keV. 
Our default energy interval is $[3, 30.5]$~keV, where the threshold is
motivated by the LUX analysis\cite{LUX:2013}, but we show also results
for $[6.5, 30.5]$~keV, corresponding to the benchmark energy interval
of the XENON100 experiment~\cite{Aprile:2012nq}. We adopt a Gaussian
energy resolution with width of $0.1 \sqrt{E_{nr}/E_{thr}}$~keV, where
$E_{thr}$ is our chosen threshold energy. This is a very optimistic
assumption on the energy resolution, however, we have checked that our
results are very insensitive to the assumed energy resolution.

\begin{table}
\centering
  \begin{tabular}{lcccccc}
  \hline\hline
  DM mass [GeV] & 20 & 50 & 80 & 100 & 200 \\
  \hline
   \# events (no GF)   & 11305 & 21094 & 16974 & 14514 & 8058 \\
   \# events (with GF) & 11525 & 21418 & 17213 & 14712 & 8162 \\
  \hline\hline
  \end{tabular}
  \caption{\label{tab:events} Number of events with and without GF in
    our assumed xenon detector in the nuclear recoil energy range
    $[3.0, 30.5]$~keV with an exposure of $270$~ton~yr for
    $\sigma_{\rm SI} = 10^{-45}$~cm$^2$ and various values of the DM
    mass.}
\end{table}

We assume now a reference value for the spin-independent cross section
of $\sigma_{\rm SI} = 10^{-45}$~cm$^2$, close to the present upper
limit. In Tab.~\ref{tab:events} we give the total number of events
predicted with and without GF for our adopted configuration. We expect
of order $10^4$ events, with some dependence on the DM mass. The
effect of GF on the total event numbers is very small (and degenerate
with the scattering cross section or other global normalization
factors, such as e.g., $\rho_\infty$).

\subsection{The annual modulation signal with and without GF}

\begin{figure}
\begin{center}
 \includegraphics[width=0.49\textwidth]{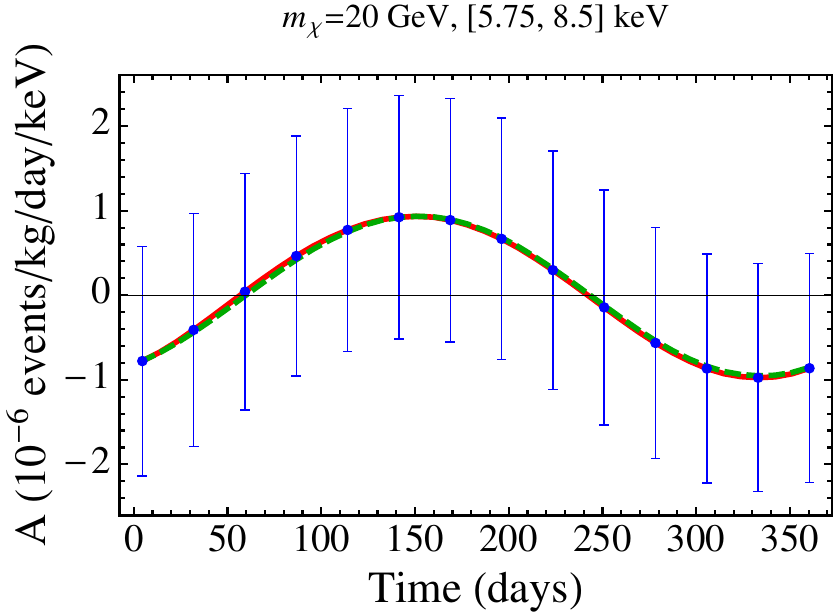}
 \includegraphics[width=0.49\textwidth]{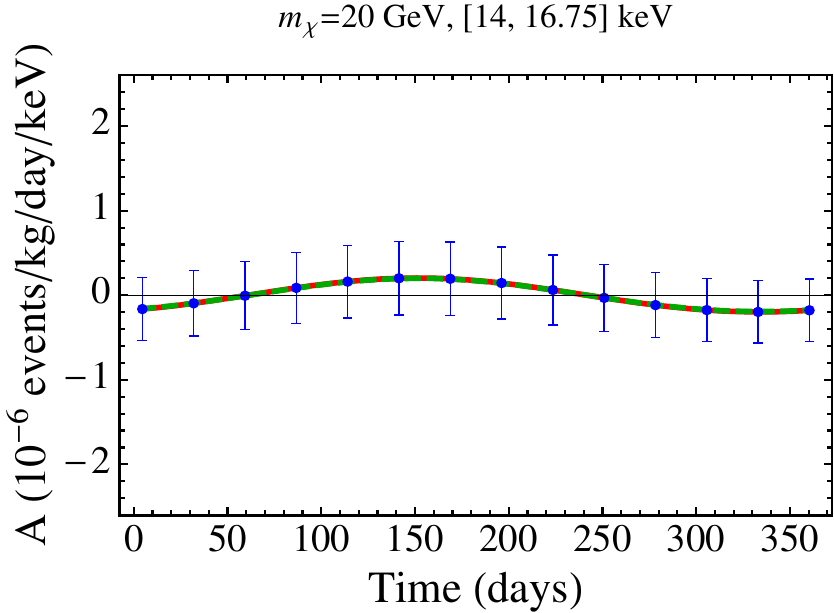}\\[5mm]
  \includegraphics[width=0.49\textwidth]{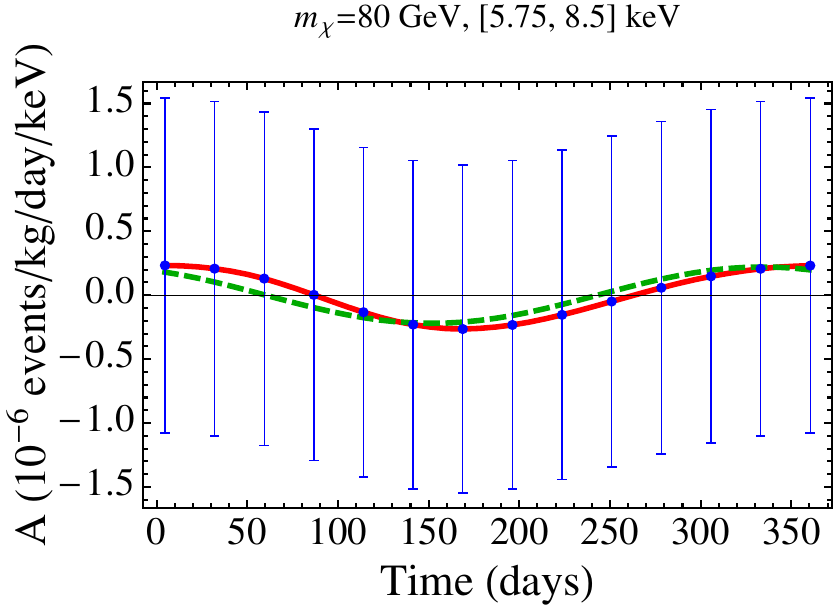}
 \includegraphics[width=0.49\textwidth]{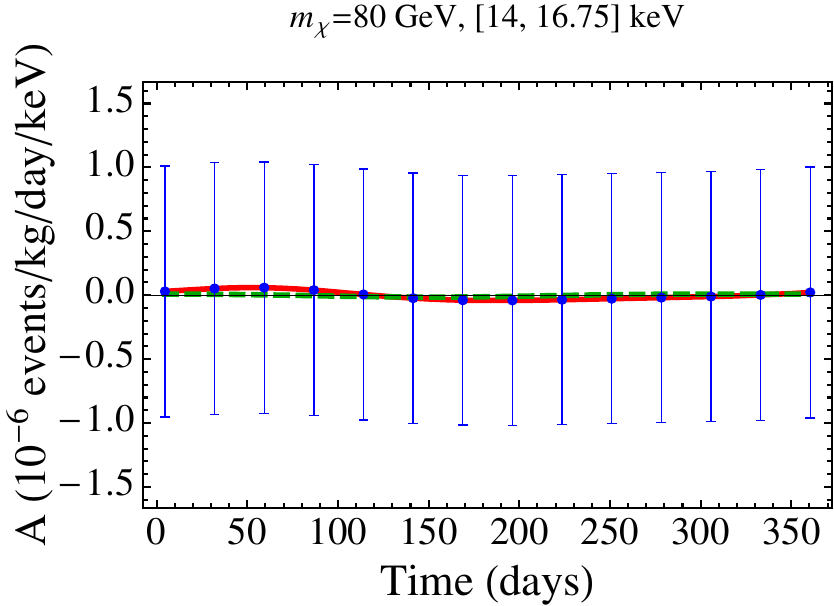}\\[5mm]
 \includegraphics[width=0.49\textwidth]{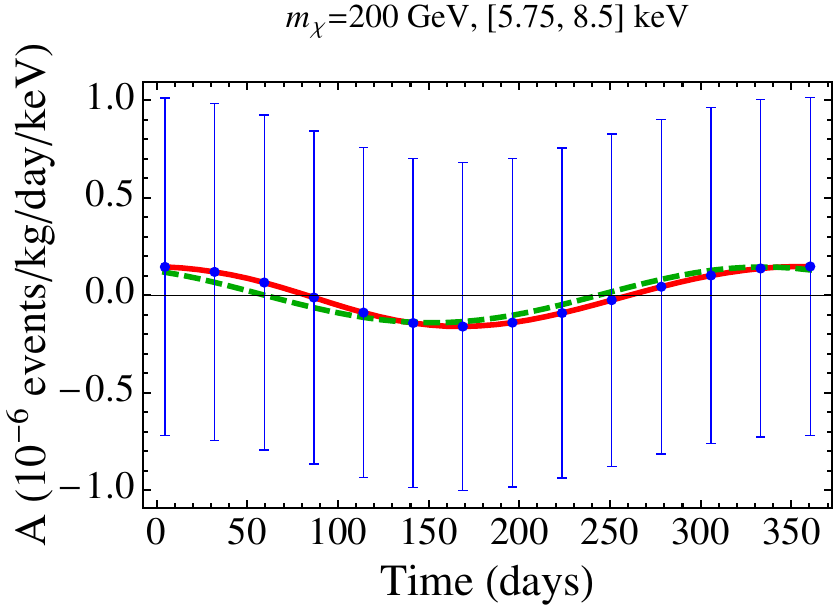}
 \includegraphics[width=0.49\textwidth]{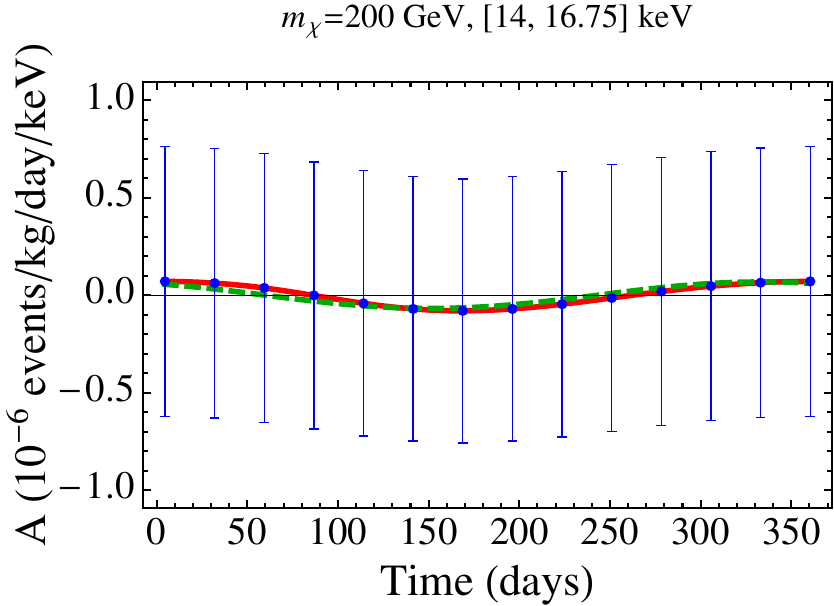}
\caption{\label{fig:Modulation} The annual modulation signal
  $A_{[E_1,E_2]}/MT$, see Eq.~\eqref{eq:A}, with GF (blue points and
  red curve) and without GF (green dashed curve) for a xenon target
  and a spin-independent cross section of $\sigma_{\rm SI} =
  10^{-45}$~cm$^2$. In the top, center, and bottom panels,
  $m_\chi=20$, 80, and 200~GeV, respectively. In the left and right
  panels, the energy interval chosen is $[5.75, 8.5]$ keV and $[14,
    16.75]$ keV, respectively. Error bars correspond to the
  statistical error in $A_{[E_1,E_2]}/MT$ including GF assuming an
  exposure of $MT= 10^8$~kg~day~$\approx 270$~ton~yr. Note the different
  scales on the vertical axes.}
\end{center}
\end{figure}

Let us now consider the size of the annual modulation signal for the setup described above.
In Fig.~\ref{fig:Modulation} we show plots of $A_{[E_1,E_2]}/MT$ (in events/kg/day/keV), where we divide the modulation amplitude given in Eq.~\eqref{eq:A} by the exposure $MT$, as a function of time for two different energy bins (of equal size). The blue points as well as the red curve show the predicted annual modulation signal with GF, and the error bars show their statistical error averaged over each time bin. The green dashed curve shows the annual modulation without GF. In the upper panels we assume a DM mass of 20~GeV. In this case the predicted signals with and without GF are identical. For such small masses we are probing large values of $v_m$ (364~km/s and  535~km/s in the center of the low and high energy bins in the left and right upper panels of Fig.~\ref{fig:Modulation}, respectively), and in that region GF is negligible, compare to Fig.~\ref{fig:eta}. Some differences in the predictions are visible in the middle and bottom panels, corresponding to DM masses of 80 and 200~GeV, respectively, and hence lower values of $v_m$ where GF becomes important. Considering the size of the error bars, it is clear that establishing GF -- or even the annual modulation itself -- at high significance will turn out to be hard. We will come back to this question below in subsection~\ref{sec:statistics}. Note that in the high-energy bin for $m_\chi = 80$~GeV (right-middle panel) the modulation amplitude is very small. In this case we are very close to the phase flip ($v_m = 190$ km/s in the center of the bin) where the amplitude is suppressed.

\begin{figure}
\begin{center}
 \includegraphics[width=0.49\textwidth]{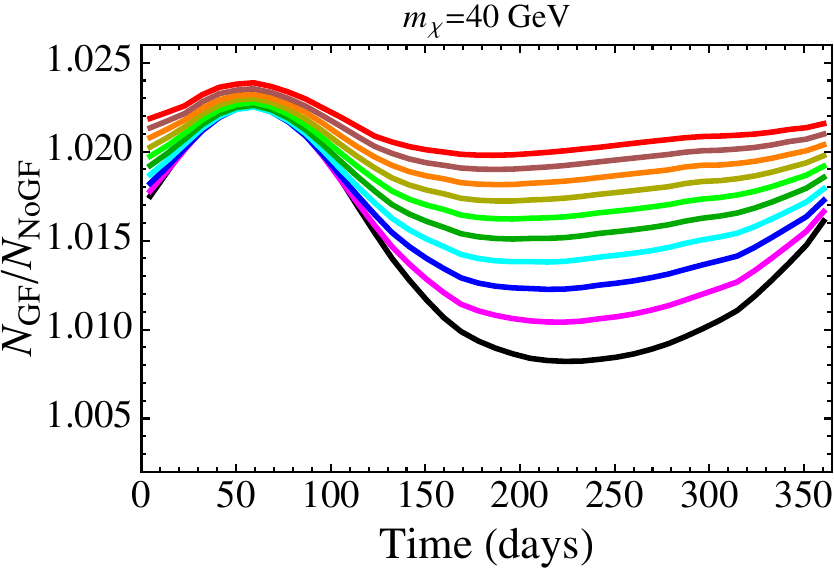}
 \includegraphics[width=0.49\textwidth]{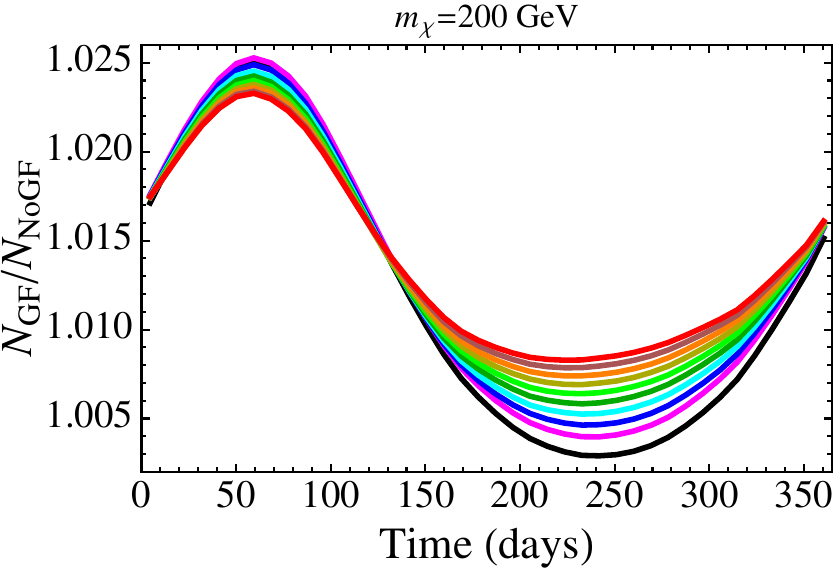}\\
\caption{\label{fig:Ratio} The ratio of the number of events with and without GF as a function of time and in different energy bins. The black to red curves correspond to energy bins of equal size in the interval of $[3.0, 30.5]$ keV from lowest to highest, respectively. The left and right panels are for $m_\chi=40$ GeV and 200 GeV, respectively. We use a Gaussian energy resolution with a width of $0.1 \sqrt{E_{nr}/3.0}$ keV.}
\end{center}
\end{figure}

Fig.~\ref{fig:Ratio} shows the ratio of the number of events with GF, $N_{\rm GF}$, and without GF, $N_{\rm NoGF}$, as a function of time and in different energy bins for the two values of the DM mass of 40~GeV and 200~GeV. The 10 colored lines correspond to 10 energy bins of equal size in the interval of $[3.0, 30.5]$ keV starting from the lowest energy bin (black) to the highest one (red). Fig.~\ref{fig:Ratio} shows that the effect of GF is smaller for $m_\chi=40$ GeV compared to $m_\chi=200$ GeV. For both DM masses, the ratio is close to 1, with the size of the effect at the percent level.

\subsection{Statistical significance of modulation and GF}
\label{sec:statistics}

We proceed now by estimating the achievable statistical significance
for the annual modulation as well as for GF obtainable by our
hypothetical xenon detector with an exposure of $MT \approx 270$~ton~yr, as
described in section~\ref{sec:exp-details}. Again we take as reference
value a cross section close to the present LUX bound of $\sigma_{\rm
  SI} = 10^{-45}$~cm$^2$.  We divide the ``data'' into 40 time bins
per year and 10 energy bins of equal size in the energy interval of
either $[3, 30.5]$~keV or $[6.5, 30.5]$~keV (in order to illustrate
the importance of the threshold energy). Then we construct a $\chi^2$
function based on the annual modulation amplitude as defined in
Eq.~\eqref{eq:A}:
\be
\Delta\chi^2 (m_\chi, \sigma_{\rm SI}; \, m_\chi^0, \sigma_{\rm SI}^0) 
= \sum_{i, j} \left(\frac{A_{ij}^{\rm pred} (m_\chi, \sigma_{\rm SI}) - 
A_{ij}^{\rm obs}(m_\chi^0, \sigma_{\rm SI}^0) }{\sigma_{ij}} \right)^2,
\label{eq:Xechisq}
\ee
where the sum is over both time and energy bins, and $\sigma_{ij}$
is the statistical error of $A_{ij}^{\rm obs}$. Here, $A_{ij}^{\rm
  obs}$ plays the role of the future ``data'', for which we take the
predicted modulation amplitude for some particular ``true'' parameter
values $(m_\chi^0, \sigma_{\rm SI}^0)$, where we fix $\sigma_{\rm
  SI}^0 = 10^{-45}$~cm$^2$ as mentioned above, but vary the assumed
true DM mass $m_\chi^0$. Note that we do not include statistical
fluctuations in our ``data'', but instead use the most probable
outcome of the experiment as data. This is sometimes called ``Asimov
data set'' \cite{Cowan:2010js} and $\chi^2$ values obtained in this
way describe the significance of the ``average'' experiment, see for instance 
\cite{Newstead:2013pea} for an application of this approach in the dark matter context.

\begin{figure}
\begin{center}
 \includegraphics[width=0.46\textwidth, height=0.37\textwidth]{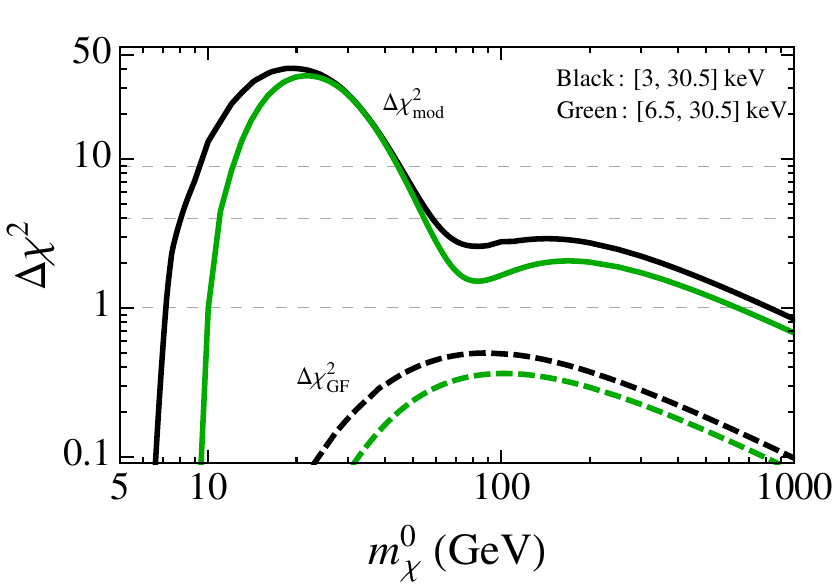}
 \includegraphics[width=0.53\textwidth, height=0.37\textwidth]{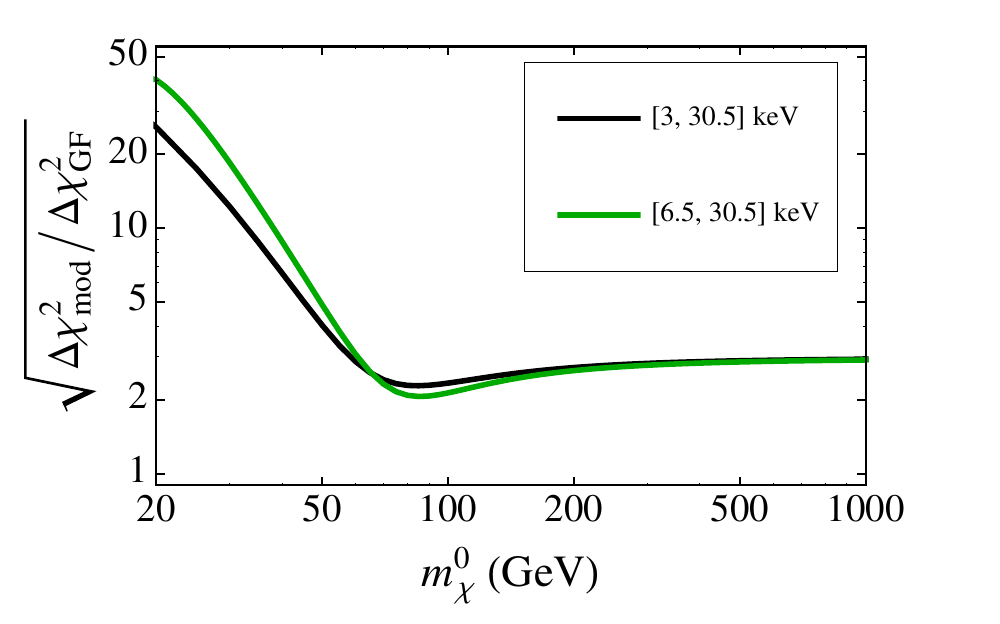}
\caption{\label{fig:chisq} Left: The $\Delta \chi^2_{\rm mod}$  (Eq.~\eqref{eq:Xechisq-mod}) is shown as a function of $m_\chi^0$ by solid lines. The dashed lines show the $\Delta \chi^2_{\rm GF}$  computed from Eq.~\eqref{eq:Xechisq} assuming the predicted signal is without GF. Right: The square root of the ratio $\Delta \chi^2_{\rm mod} / \Delta \chi^2_{\rm GF}$ as a function of $m_\chi^0$. We use an energy threshold of 3.0 keV and 6.5 keV for the black and green curves, respectively. We assume $\sigma_{\rm SI}=10^{-45}$ cm$^2$ and an exposure of $10^8$ kg day~$\approx 270$~ton yr.}
\end{center}
\end{figure}

First we estimate the significance with which the presence of annual
modulation can be established. To this aim we calculate the ``data''
$A_{ij}^{\rm obs}$ as described above for a choice of $m_\chi^0$ and
our reference cross section, including GF. The $\chi^2$ value obtained
then by fitting these data with a prediction constant in time will
measure the significance of the modulation. Hence we set $A_{ij}^{\rm
  pred} = 0$ in Eq.~\eqref{eq:Xechisq}:
\be
\Delta\chi^2_{\rm mod} (m_\chi^0, \sigma_{\rm SI}^0) 
= \sum_{i, j} \left(\frac{A_{ij}^{\rm obs}(m_\chi^0, \sigma_{\rm SI}^0) }{\sigma_{ij}} \right)^2 \,.
\label{eq:Xechisq-mod}
\ee
Evaluating $\Delta\chi^2_{\rm mod} (m_\chi^0, \sigma_{\rm SI}^0)$
for 1 degree of freedom corresponds to the average significance for
annual modulation, and hence, $\sqrt{\Delta\chi^2_{\rm mod} (m_\chi^0,
  \sigma_{\rm SI}^0)}$ gives the corresponding number of Gaussian
standard deviations. The result of this analysis is shown with solid
curves in the left panel of Fig.~\ref{fig:chisq} for two different
choices of energy threshold. For the black curve we use the energy
interval $[3, 30.5]$~keV, whereas for the green curve we use $[6.5,
  30.5]$~keV. We conclude from this figure that only in the mass range
$m_\chi \lesssim 40$~GeV a significant ($> 3\sigma$) signal for annual
modulation can be obtained, whereas for larger DM masses only a hint
below $2\sigma$ can be reached, despite the huge exposure we are
assuming.

Next we want to estimate the significance of gravitational
focusing. We formulate this in terms of a hypothesis test, where the
presence of GF is the null hypothesis which we want to test against
the alternative hypothesis of the absence of GF. Given data, one would
perform a fit with and without GF and compute a test statistic $T$,
for instance $T = \chi^2_\text{with GF} - \chi^2_\text{without
  GF}$. One can show that in the Gaussian approximation $T$ will be 
normal distributed with mean $T_0$ and standard deviation
$2\sqrt{T_0}$ with $T_0 = \Delta\chi^2_{\rm GF}$, where $\Delta
\chi^2_{\rm GF}$ is obtained from Eq.~\eqref{eq:Xechisq} by
calculating $A_{ij}^{\rm obs}$ including GF and the
predicted signal, $A_{ij}^{\rm pred}$, without GF. A detailed
discussion of the statistical method is given in
Ref.~\cite{Blennow:2013oma}, where a similar method is applied to the
problem of identifying the neutrino mass ordering. In that reference
also a derivation of the normal distribution of $T$ can be found.
Following Ref.~\cite{Blennow:2013oma}, we note that
$\sqrt{\Delta\chi^2_{\rm GF}}$ corresponds to good approximation to
the number of standard deviations with which the median experiment can
reject the alternative hypothesis. The dashed curves in the left
panel of Fig.~\ref{fig:chisq} show the $\Delta \chi^2_{\rm GF}$ as a
function of $m_\chi^0$.\footnote{Note that for given true values
  $(m_\chi^0,\sigma_{\rm SI}^0)$, $\Delta \chi^2_{\rm GF}$ would still
  depend on the fitted $(m_\chi,\sigma_{\rm SI})$ which should be
  minimized over.  We have checked that for the low energy threshold
  of 3~keV, the minimum $\Delta \chi^2_{\rm GF}$ values differ only by
  9\% at $m_\chi^0 = 20$~GeV, and less than 0.5\% for DM masses above
  80~GeV compared to evaluating both, $A_{ij}^{\rm pred}$ and
  $A_{ij}^{\rm obs}$ at the true parameter values, i.e., fixing
  $m_\chi = m_\chi^0$ and $\sigma_{\rm SI} =\sigma_{\rm SI}^0$. In our
  results we adopt this approximation since it implies a huge saving
  in computation time.}  The black and green curves are for an energy
threshold of 3 keV and 6.5 keV, respectively. We conclude that for our
assumed setup the significance of detecting gravitational focusing
never reaches the 1$\sigma$ level.

Note that our analysis is based on statistical errors only. This
implies that all our $\chi^2$ values are proportional to the overall
event normalization, and therefore scale linearly with the exposure
$MT$ times the scattering cross section $\sigma_{\rm SI}^0$. Hence the
curves in the left panel of Fig.~\ref{fig:chisq} can easily be
translated to any other exposure and/or cross section. In view of this
observation we show in the right panel of Fig.~\ref{fig:chisq} the
quantity $\xi \equiv \sqrt{\Delta\chi^2_{\rm mod} / \Delta \chi^2_{\rm
    GF}}$. This number can be interpreted in the following way: assume
that a future experiment established annual modulation at $n\sigma$
significance; then GF will be detected at a significance of
$(n/\xi)\sigma$. The ratio $n/\xi$ is largest when $\Delta
\chi^2_{\rm GF}$ reaches a maximum and $\Delta \chi^2_{\rm mod}$
reaches a local minimum, which happen at $m_\chi^0 \sim 70 - 90$~GeV
depending on the choice of energy threshold. This behavior is easy
to understand, since as mentioned above, for those masses the phase
flip of the modulation is located in the middle of the energy
spectrum. This minimizes the amplitude of the modulation and explains
the local minimum in the $\chi^2_{\rm mod}$ in the left panel (see
also right-middle panel in Fig.~\ref{fig:Modulation}), whereas around
the phase flip the impact of GF on the phase of the modulation is
strongest (see Fig.~\ref{fig:eta}).

One might ask if the results are dependent on the choice of the number of time or energy bins in the calculation of $\Delta \chi^2$. We have checked that the significance of detecting the annual modulation and GF stay the same for various choices of the number of  time and energy bins, as long as they are larger than a few. In particular, we observe that the energy spectrum seems not to be very important for our results. Thus the choice of 10 energy bins and 40 time bins give accurate results. This conclusion may change in the presence of systematic uncertainties and/or backgrounds, since in this case the particular energy shape of the signal may be important to limit the impact of such uncertainties. We have also checked that our choice of energy resolution is not crucial for our results.

\begin{figure}
\begin{center}
 \includegraphics[width=0.50\textwidth]{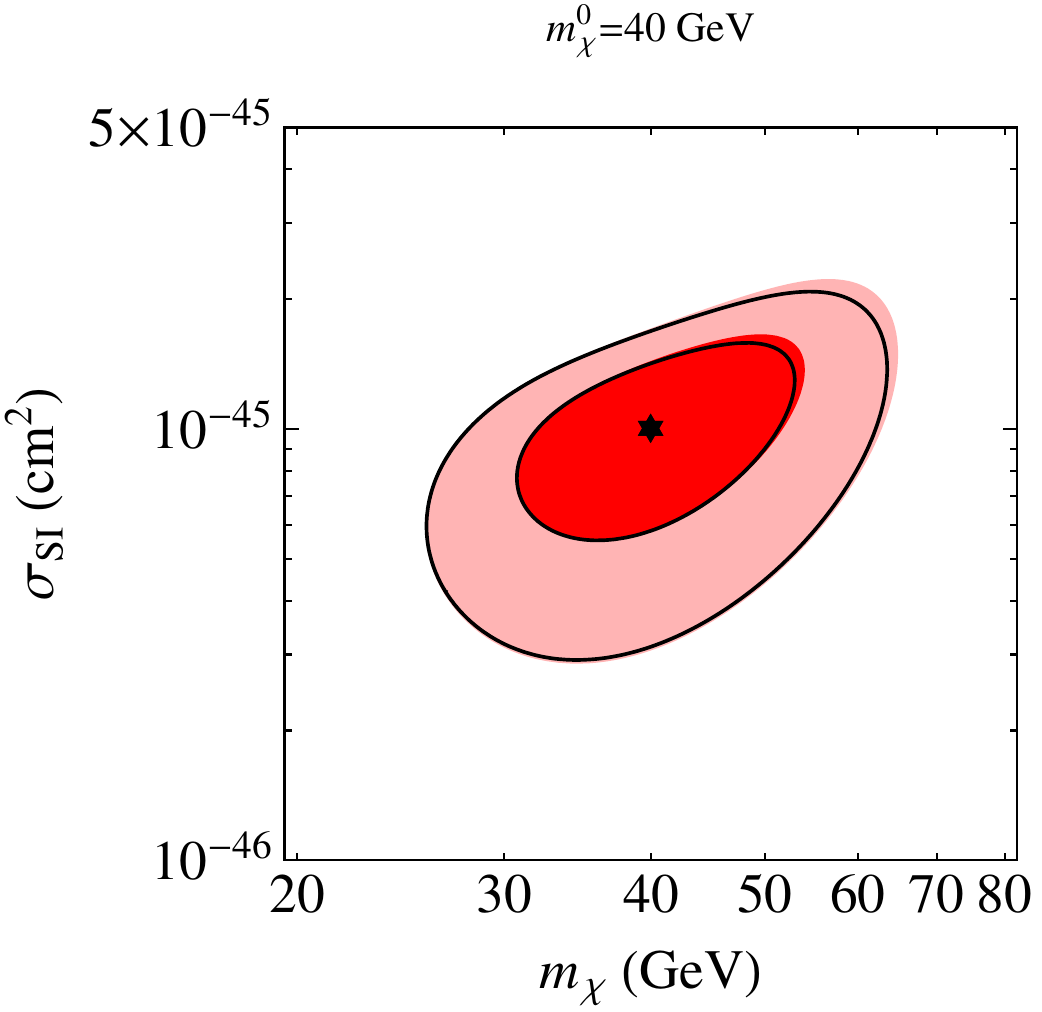}\hspace{2pt}
 \includegraphics[width=0.48\textwidth]{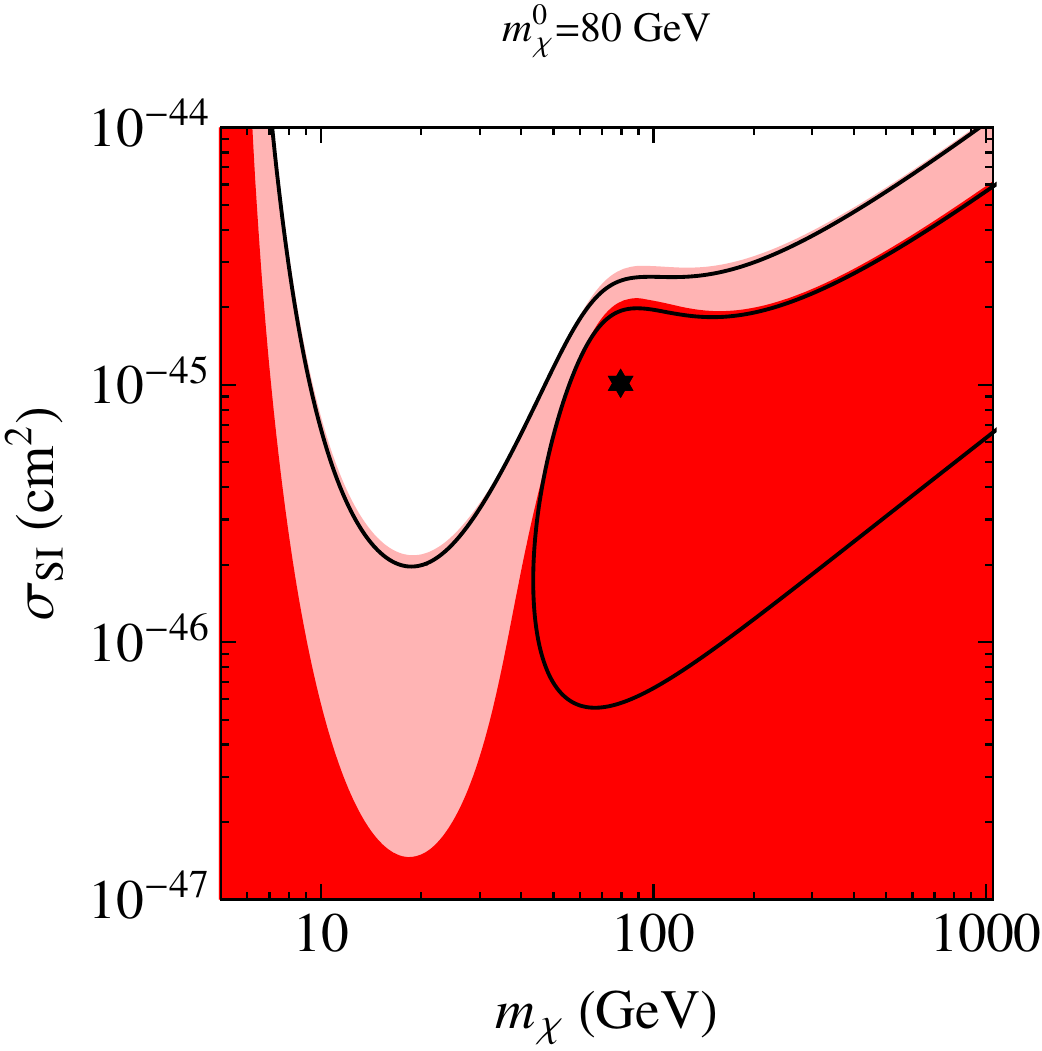}
\caption{\label{fig:Xe} The preferred regions at 1$\sigma$ and
  2$\sigma$ based on the annual modulation signal in a future xenon
  experiment with an exposure of 270~ton~yr with GF (black contours) and without GF
  (dark and light red regions). In the left and right panels, we
  assume a positive signal at $m_\chi^0=40$~GeV, and 80~GeV,
  respectively, both with a cross section of $10^{-45}$ cm$^2$, as
  specified with a star in the plots.}
\end{center}
\end{figure}

To conclude this section we give two examples of the preferred regions
in the DM mass--cross section plane obtainable from the annual
modulation signal in our hypothetical future xenon detector. We
calculate the ``data'' $A^{\rm obs}$ in Eq.~\eqref{eq:Xechisq} for two
example values of the DM mass, 40 and 80~GeV, and the reference cross
section $10^{-45}$~cm$^2$. Allowed regions can then be obtained by
considering contours of $\Delta \chi^2$ from Eq.~\eqref{eq:Xechisq} in
the $(m_\chi, \sigma_{\rm SI})$ plane, as shown in
Fig.~\ref{fig:Xe}. The ``data'' $A^{\rm obs}$ has been calculated
including GF. In the figure we show the allowed regions for either
including GF when calculating the prediction (black contour curves) or
neglecting GF (colored regions). In the first case the $\chi^2$
minimum at the true parameter point is zero, whereas in the second
case the minimum $\chi^2$ is non-zero (though very small, compare to
Fig.~\ref{fig:chisq}). In this case contour levels are defined with
respect to the (non-zero) minimum.  As it can be seen from
Fig.~\ref{fig:Xe}, GF affects only marginally the allowed region at
40~GeV, as expected from Fig.~\ref{fig:chisq} and the observation that
large $v_m$ values are probed for small DM masses. The difference for
the $1\sigma$ region at 80~GeV looks striking, being closed from below (open)
with (without) GF. There is no need to emphasize, however, that $1\sigma$ regions in
general are not statistically meaningful, and already at $2\sigma$
the regions with and without GF are very similar.

Let us stress that here we discard any information on the absolute
number of events and base the regions only on the annual modulation
signal. Clearly, with event numbers of order $10^4$ in our setup (see
Tab.~\ref{tab:events}) in principle a very precise mass and cross
section determination will be possible, assuming that backgrounds as well as
astrophysical uncertainties are under control, see e.g.,
Ref.~\cite{Newstead:2013pea} for analyses along those lines in the
context of DARWIN. The purpose of the discussion here is to
investigate the potential (or the lack thereof) of the annual
modulation signal, which is supposed to be more robust against
systematic uncertainties.

\subsection{Comments on light targets}
\label{sec:light}

Let us briefly comment also on lighter target nuclei and experiments
aiming at very low threshold, such as for instance the CDMSlite
experiment~\cite{Agnese:2013jaa}. In this case in principle somewhat
lower $v_m$ values are probed even for light DM, where the modulation
amplitude is larger. Therefore, a valid question is whether GF becomes
more important for such configurations. We have performed similar
tests as the one described above for a Xe detector also for setups
using Ar and Ge. In general the signal is reduced compared to Xe (for
the same nucleon cross section and exposure) due to the $A^2$
dependence of the rate. However, the shape of the curves for
$\Delta\chi^2_{\rm mod}$ and $\Delta\chi^2_{\rm GF}$ as a function of
$m_\chi$ are similar to the case of xenon. In particular, for an argon
detector (with threshold 3~keV) we obtain a very similar result for
the ratio shown in the right panel of Fig.~\ref{fig:chisq}.

The germanium based experiment CDMSlite~\cite{Agnese:2013jaa} has
achieved a threshold of 0.84~keV. This opens the
possibility to test very low DM masses, where the modulation signal
can be very significant. We tested a hypothetical Ge detector with a 
threshold of 0.84~keV and again we arrive at qualitatively similar
conclusions as for Xe. For example, for a DM mass of 4~GeV we can take
a cross section of $10^{-40}$~cm$^2$, just below the current limits. In
this case we would need an exposure of about $10^6$~kg~day in order to
see a $3\sigma$ effect of GF ($\Delta \chi^2_{\rm GF} \sim 10$). This
assumes a background free experiment and has to be compared to the
current CDMSlite exposure of about 6~kg~day~\cite{Agnese:2013jaa}.

\subsection{Comments on inelastic scattering}
\label{sec:inelast}

Above we focused exclusively on elastic spin-independent
scattering. Let us comment briefly on other particle physics
scenarios. We do not expect any qualitative change for other
DM--nucleus interactions, which provide some kind of re-scaling of the
scattering rate, such as spin-dependent scattering or couplings of
different strength to neutrons and protons. There may be, however,
non-trivial effects for interaction types which modify the kinematics
and/or change the ratio of the amplitude of the annual modulation to
the time-constant signal. One example of this type is inelastic
scattering, $\chi + A \to \chi' + A$, with the mass difference $\delta
= m_{\chi'} - m_\chi$. In this case the expression for the minimal
velocity, Eq.~\eqref{eq:vm}, is changed to
\beq\label{eq:vm-inel}
v_m=\frac{1}{\sqrt{2 m_A E_{nr}}} \left | \frac{m_A E_{nr}}{\mu_{\chi A}} + \delta \right |.
\eeq

In the case of endothermic scattering, $\delta>0$, the DM particle
up-scatters to an excited state~\cite{TuckerSmith:2001hy}. In this
case scattering off heavier target nuclei is favored.  It is clear
from Eq.~\eqref{eq:vm-inel} that for $\delta > 0$, $v_m$ gets larger
compared to the elastic case, and hence, we expect that GF becomes
less important (c.f. Fig.~\ref{fig:eta}). Indeed, we find that for
endothermic scattering the significance of detecting GF is very small,
and $\Delta \chi^2_{\rm GF}$ never reaches even 0.1 for our setup.

\begin{figure}
\begin{center}
 \includegraphics[width=0.49\textwidth, height=0.37\textwidth]{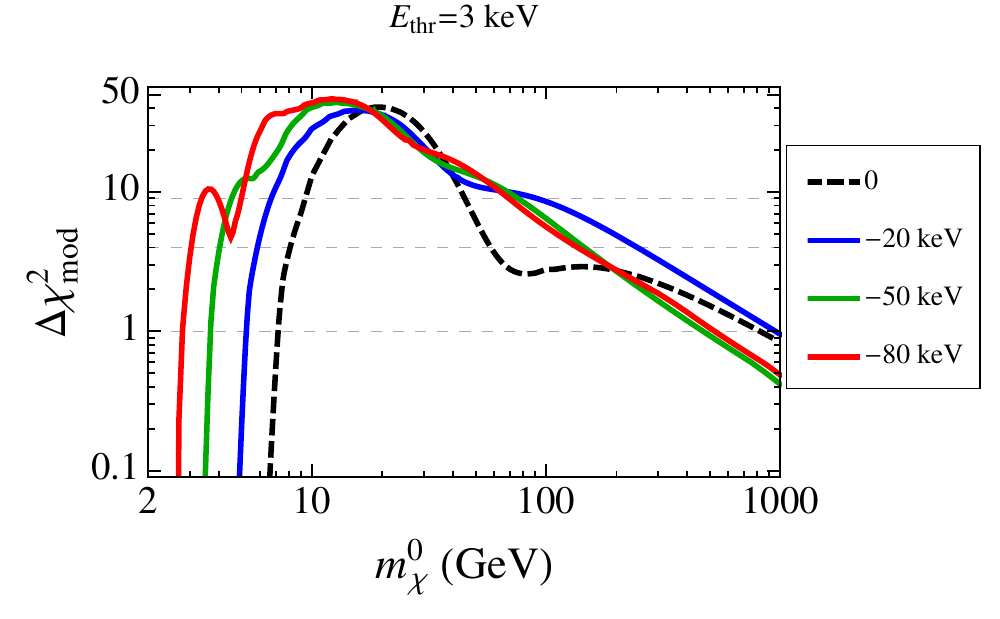}
 \includegraphics[width=0.49\textwidth, height=0.37\textwidth]{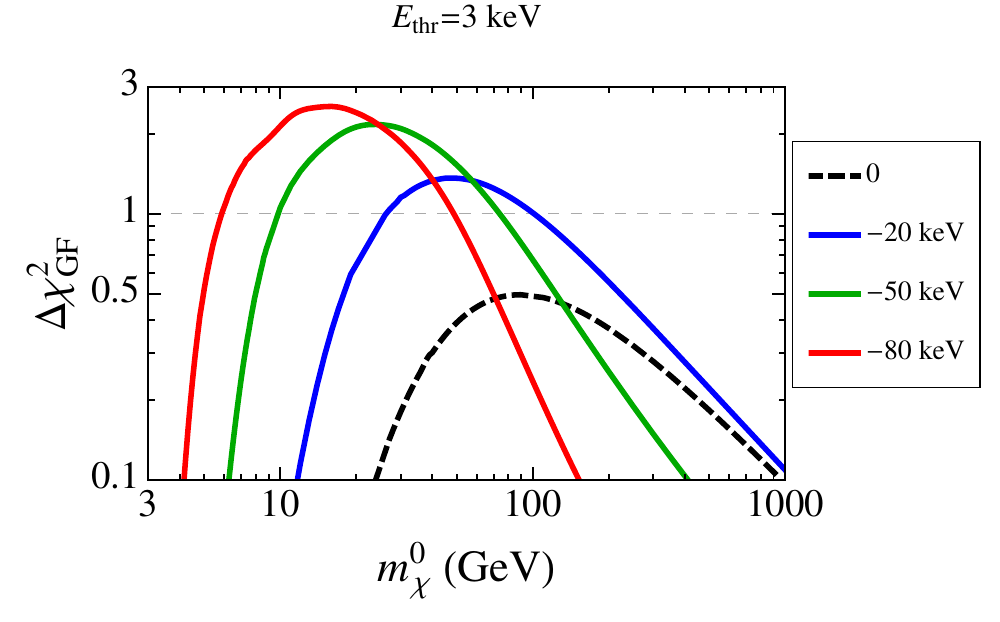}
\caption{\label{fig:chisq-IE} The $\Delta \chi^2_{\rm mod}$ (left) and
  the $\Delta \chi^2_{\rm GF}$ (right) for inelastic exothermic
  scattering, shown as a function of $m_\chi^0$ using an energy
  threshold of 3.0 keV. The dashed black curve corresponds to the elastic
  case ($\delta=0$), while the other colors correspond to different
  values of $\delta < 0$. We assume $\sigma_{\rm SI}=10^{-45}$ cm$^2$
  and a Xe exposure of $270$~ton yr.}
\end{center}
\end{figure}

On the other hand, if $\delta<0$, the DM particle down-scatters to a
lower mass state, which is called exothermic
scattering~\cite{Finkbeiner:2007,Batell:2009,Graham:2010}, and in this
case we are probing the region of smaller $v_m$ where GF can be
important. The significance of detecting the annual modulation and GF
are shown in the left and right panels of Fig.~\ref{fig:chisq-IE},
respectively, for different choices of negative $\delta$.  Here we
again assume a Xe detector as described in
section~\ref{sec:exp-details}. The case of elastic scattering
($\delta=0$) is also shown in dashed black. As one can see from the right
panel of Fig.~\ref{fig:chisq-IE}, the significance of detecting GF can
reach values above 1$\sigma$ for $|\delta| \gtrsim 20$~keV.

The wiggles at low masses in the $\Delta \chi^2_{\rm mod}$ curves for negative $\delta$ (most pronounced for $\delta=-80$ keV) occur when $v_m$ goes through zero for different energy bins, see Eq.~\eqref{eq:vm-inel} for $\delta < 0$. For the WIMP mass and recoil energy at which $v_m=0$, we are integrating the velocity distribution over a large range of velocities, which leads to a small amplitude for the annual modulation. Thus for such masses, $\Delta \chi^2_{\rm mod}$ goes through a local minimum and the significance of detecting the annual modulation is smaller.

\section{Summary}
\label{sec:summary}

In this paper we have considered the impact of the Sun's gravitational
potential on the annual modulation signal in dark matter direct
detection experiments.  The distortion of the DM phase space
distribution due to the gravitational focusing (GF) of the Sun may
potentially lead to a significant modification of the time dependence
of the count rate \cite{Lee:2013wza}. To illustrate the importance of
this effect we have considered the signal reported by the DAMA/LIBRA
experiment. Performing a fit to their time dependent data we find that
the allowed DM parameter range is more restricted (for large DM
masses) if the effect of GF is taken into account. Since, however, for
those DM masses the relevant cross section is excluded by many orders
of magnitude by recent constraints from other experiments, we set the
DAMA/LIBRA signal aside and investigate the potential of possible
future large-scale direct detection experiments.

We consider a very large xenon-based setup, with an exposure around
$10^4$ times larger than the current exposure from the LUX experiment
\cite{LUX:2013}, corresponding to about 270~ton~yr. Furthermore, we
assume that DM is just around the corner, with a DM--nucleus cross
section of $\sigma_{\rm SI} = 10^{-45}$~cm$^2$, roughly at the present
LUX exclusion limit. Even under those very optimistic assumptions our
results indicate that most likely the answer to the question posed in
the title of the paper is ``no''. We find that an annual modulation
signal can be established at a significance of $\gtrsim 3\sigma$ only
for DM masses $m_\chi \lesssim 40$~GeV. For such small masses, only
the high-velocity tail of the DM distribution is probed, where the
effect of GF is very small. In the region of larger DM masses, where
GF may be potentially observable, the annual modulation signal itself
will not become significant. We have considered also the case of
inelastic scattering, where for the exothermic case (down-scattering)
the effect of GF is slightly larger than for the elastic case, because
of the lower DM velocities involved. We have also checked that our
conclusions hold in case of lighter target nuclei such as Ar and Ge.

Our calculations are based on the Standard Halo Model, corresponding
to an isotropic Maxwellian velocity distribution. It is well known
that deviations from this halo may have important consequences for the
annual modulation signal \cite{Fornengo:2003fm, Green:2003yh,
  Freese:2012xd}. Investigating the impact of GF in the presence of
more complicated DM velocity distributions, such as streams or debris
flows is beyond the scope of this work. Generically we do not expect
that the size of the GF effect will change, however, we cannot exclude
the possibility that particular configurations, for instance a dark disk, may enhance the
importance of GF. It might be a hard task to disentangle GF
effects from non-standard halos. In general, we conclude by noting
that for a given halo model the impact of GF is determined and
calculable, and in case of doubt should be included in the analysis.

\subsection*{Acknowledgements}

N.B.\ acknowledges discussions with Graciela Gelmini about
investigating the impact of gravitational focusing on the DAMA regions
during the TAUP 2013 conference. We thank the authors of
Ref.~\cite{Lee:2013wza} for valuable comments on the first version of
this paper. We acknowledge support from the European Union FP7 ITN INVISIBLES
(Marie Curie Actions, PITN-GA-2011-289442). N.B.\ thanks the Oskar
Klein Centre and the CoPS group at the University of Stockholm for
hospitality during her long-term visit.

\bibliographystyle{my-h-physrev.bst}
\bibliography{./refs}

\end{document}